\documentclass[%
 reprint,
groupedaddress,
showpacs,
 amsmath,amssymb,
prb,
]{revtex4-1}

\usepackage[final]{graphicx}
\usepackage{dcolumn}
\usepackage{bm}

\begin{document}


\title{Elastic Properties of Functionalized Carbon Nanotubes 
}

\author{Karolina Z. Milowska}
\author{Jacek A. Majewski}%
\affiliation{%
Institute of Theoretical Physics, Faculty of Physics, University of Warsaw, ul. Hoza 69, PL-00-681 Warszawa, Poland\\
}%


\date{\today}

\begin{abstract}
We study the effects of covalent functionalization of  single wall carbon nanotubes (CNT) on their elastic properties. We consider simple organic molecules -NH, -NH$_2$, -CH$_2$, -CH$_3$, -OH attached to CNTs' surface at various densities.  The studies are based on the first principles calculations in the framework of density functional theory. We have determined the changes in the geometry and the elastic moduli of the functionalized CNTs as a function of the density of adsorbed molecules. It turns out that elastic moduli diminish with increasing concentration of adsorbents, however, the functionalized CNTs remain strong enough to be suitable for reinforcement of composites. The strongest effect is observed for CNTs functionalized with  -CH$_2$  radical, where the Young's modulus of the functionalized system is by 30\%  smaller than in the pristine CNTs. 

\end{abstract}

\maketitle

\section{Introduction}
Since discovery in 1991 \cite{Iijima1991}, carbon nanotubes (CNTs)  have quickly developed as the working horse of nanotechnology, mostly owing to their remarkable electronic, mechanical and thermal properties, which facilitate a whole plethora of CNTs' applications.  Among them are new composite materials synthesized by adding CNTs to various materials such as alloys, polymers, and metals. Such composites constitute the extraordinary class of materials being very light and exhibiting simultaneously  enhanced mechanical strength, electrical and thermal conductivity, and chemical stability \cite{Terrones2003,book1}. However, the fabrication of such nano-composites is hindered by the fact that the pristine CNTs are not soluble in water or in organic solvents and have tendencies to create bundles. The common remedy of these problems is the functionalization of the CNTs, in particular covalent functionalization with simple organic molecules  (such as -CH$_n$, -NH$_n$ fragments, and -OH  groups). These molecules adsorbed at the surface of CNTs allow for strong binding of the functionalized in such a way CNTs with matrix material, typically a polymer or a metal \cite{book1, amr2011, steiner2012, lachman2010, gojny2005}. 

On the other hand, functionalization to the side walls of CNTs changes their morphology, generates defects \cite{app1, prof1, diamond, condmat}, and could decrease the strength of the structure in comparison to the pristine CNTs. Therefore, it is very important to investigate the elastic properties of the functionalized CNTs. It is also meaningful having in mind broad area of CNT applications in fields such as nanoelectronics or medicine. Generally, the elastic properties of the functionalized CNTs are rather poorly known, in contrast to the pristine ones\cite{lier2000, li2003, hernandez1998, govindjee1999, chang2006, yao1998, xin2000, Terrones2003, kudin2001, sanchez1999, lu1997, popov2000, krishna1998, shokrieh2010}. To close this gap, we have undertaken extensive and systematic {\sl ab initio} studies of elastic properties of the functionalized CNTs. The stability of the functionalized CNTs has been also studied previously in a series of publications.\cite{Li2004, pup,  Rosi2007, Shirvani, Veloso2006, wang, Strano, app1, diamond, prof1, condmat}. 
 	
	In this paper, we consider prototypes of the CNTs, namely  single wall (9,0), (10,0) and (11,0) CNTs, covalently functionalized with simple organic fragments -NH , -NH$_2$ , -CH$_2$ , -CH$_3$  and -OH. The molecules at various concentrations are attached to the side walls of CNTs being evenly distributed over the CNT surfaces. For the functionalized systems, we calculate first their equilibrium geometry and further their elastic moduli.  
	
\begin{figure}
\includegraphics[width=0.45\textwidth]{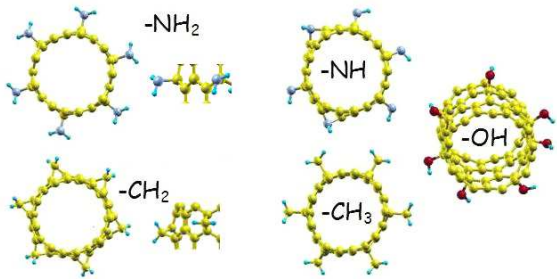}
\caption{\label{fig:Fig1} (color online) Exemplary structures of functionalized CNTs with simple -NH, -NH$_2$, -CH$_2$, -CH$_3$, and -OH  organic fragments. For -NH$_2$, -CH$_2$ fragments in addition to the cross-sectional view also the top views of local arrangement of adsorbents and surrounding C atoms are presented. This shows that depending on the electronic configuration the fragments form either single chemical bond to the C atom from the CNT backbone (-NH$_2$, -OH, and -CH$_3$), being in the so-called top position, or double bond (as -CH$_2$ and -NH) with fragment's C atom placed in the bridge positions.}
\end{figure}

		The paper is organized as follows. In Section ~\ref{sec:det}, we present calculation details.   The results of the calculations are described and discussed in the third section - 'Results and discussion'. Here we present:  (i) how the functionalization procedure changes the equilibrium geometry of the functionalized systems, (ii) how the elastic moduli of the covalently functionalized CNTs deviate from the elastic moduli of the pristine ones, and (iii) how these deviations depend on the concentration of the functionalizing molecules. Finally, the paper is concluded in section 'Conclusions'.   
We consider three types of CNTs: nominally metallic (9,0), and semiconducting (10,0) and (11,0). All of the CNTs have been covalently functionalized by attaching to their lateral surface simple organic groups, such as -NH, -NH$_2$, -CH$_2$, -CH$_3$,  and -OH. We examine those systems at various concentrations reaching up to 4.6$\cdot$10$^{14}$ adsorbed molecules per cm$^2$ of CNT's surface (see Fig.~\ref{fig:Fig1}).
However, in the present paper, we follow convention of the other authors and measure the concentration of the adsorbents as the number of attached molecules $n_A$ per doubled unit cell of pristine CNTs, i.e., per number of carbon atoms in the doubled unit cell $n_{cell}$ equal to 72, 80, and 88 carbon atoms for  (9,0), (10,0), and (11,0) CNTs, respectively. To facilitate comparison between different CNTs, we express also the concentration of the adsorbed molecules as $\frac{n_A}{n_{cell}} 100\% $. We have considered all possible positions of the adsorbed fragments and determined these positions that lead to the minimum of the total energy of the functionalized CNTs 11, i.e., the equilibrium geometry. Only these positions are depicted in Fig.~\ref{fig:Fig1}. 

The total energies and components of stress tensor are obtained from the \textit{ab initio} calculations  in the framework of the density functional theory \cite{Hohenberg1965,kohn} employing the following realization. We use the generalized gradient approximation (GGA) for the exchange-correlation density functional \cite{Perdew1996} and supercell geometry within the numerical package SIESTA \cite{siesta1, siesta2}. Since in many cases we have to do with system with odd number of electrons, we employ spin-polarized version of the GGA functional. A kinetic energy cut-off (parameter MeshCutoff in the SIESTA code) of 300 Ry and split double zeta basis set with spin polarization have been used in all calculations. Each supercell contains two primitive unit cells along the CNT symmetry axis. The lateral separation (i.e., lateral lattice constants in the direction perpendicular to the symmetry axis) has been set to 30 \AA, just to eliminate completely the spurious interaction between neighboring cells. We use the self-consistency mixing rate of 0.1, the convergence criterion for the density matrix of 10$^5$, maximum force tolerance equal to 0.01 eV/\AA, and  1x1x10 k-sampling in Monkhorst Pack scheme.

The stability of the functionalized structures can be assessed by considering the adsorption energy $E_{ads}$ defined below (and sometimes called packing energy \cite{Coto2011}). 

\begin{equation}
\label{eee}
E_{ads}  = \frac{1}{N} ( E_{CNT + groups}  - (E_{CNT}  + N \cdot E_{group} )),
\end{equation}	
where $E_{CNT + groups}$, $E_{CNT}$, and $E_{group}$ are the total energies of the functionalized CNTs with the optimized unit cell lengths and atomic geometry, pristine one, and one functionalizing molecule, respectively.

The functionalized CNTs change their lattice constants $l$ along the symmetry axis and radii $r$ in comparison to pristine ones. These parameters take the values that minimize $E_{CNT + groups}$ with the optimized positions of all atoms in the supercell (i.e., with vanishing all forces on atoms). Since the cross-sections of the functionalized CNTs in the plane perpendicular to the CNT symmetry axis, are not circular any more, we determine average radius $r$ of the functionalized nanotubes as geometrical average of carbon atom positions on the CNT surface.

Since we use the numerical code with the localized basis, the basis set superposition error (BSSE) correction should be taken into account. We have calculated this correction following the well established procedure \cite{bsse, bsse1, bsse2}
\begin{equation}
E_{cc}  = \frac{1}{N} \left( E_{CNT + \cdot ghost}  - E_{CNT}  + E_{ghost + groups} - N \cdot E_{group} )\right),
\label{eq:bsse} 
\end{equation}
where $E_{CNT +  ghost}$ and $E_{ghost + groups}$ are Kohn-Sham energies of the functionalized system but where the adsorbents or nanotube are replaced  by their ghosts \cite{siesta2}, respectively. These calculations have been performed with atomic sites fixed to their equilibrium positions. $E_{cc}$ corrects values of $E_{ads}$  by approximately 11$\%$ to 44 $\%$ and does not change the conclusions about stability of functionalized systems\cite{condmat}. We would like to stress that the BSSE correction to the adsorption energy originates mostly from the calculation of the total energies of free groups $E_{group}$.  These energies when calculated with the basis functions connected to few atoms differ considerably from energies calculated employing full basis of the whole functionalized system. The role of BSSE correction gets completely negligible when one calculates equilibrium geometry of the functionalized systems or elastic properties, since these quantities are determined on the basis of total energies for the whole functionalized systems where the bases are identical (up to atomic positions).  
This has been confirmed by calculating $dE_{cc}$/$dl$ according to the procedure described in the Ref.\cite{bsse2}.

Having determined equilibrium geometry of the functionalized CNTs, we are in the position to calculate their elastic moduli. To do so, we strain (usually we apply tensile strain) the functionalized CNTs along the symmetry axis by $\Delta l$ and calculate the response. 

The most interesting quantity, Young's modulus, has been determined in two ways: 

(i) - by comparing the total energy of unstrained ($E_l$) and strained ($E_{l+\Delta l}$) systems

\begin{equation}
\label{e1}
\nonumber Y = \frac{1}{{V_o }}\frac{{\partial ^2 E_{strain} }}{{\partial \varepsilon _{ii} ^2 }},
\quad
 E_{strain}  = E_{l+\Delta l }  - E_l, \\
\quad
 \nonumber \varepsilon_{ii}=\frac{\Delta l}{l},
\end{equation}
 where $l$ is a lattice constant along the axis of the functionalized tube, $\Delta l$ is elongation in the chosen direction, and $V_o$ is volume of the unstressed system, and (ii)  from components ($\sigma_{ii}$) of the stress tensor $Y = \sigma_{ii} / \varepsilon_{ii} $.
Volume of the pure CNT has been calculated using following relation $V_o  = 2 \cdot \pi  \cdot r \cdot l \cdot t$,
where thickness $t$ has been chosen as double Van der Waals radius of C atom (equal to 0.34 nm) \cite{lu1997,hernandez1998, sanchez1999, harik2002, li2003, chang2006}. In the case of functionalized CNT, we neglect volume of the attached molecules.

Bulk modulus and Shear modulus have been also calculated according to the formulas:
\begin{equation}
\label{e5}
K  =  \frac{Y}{3(1-2 \nu)},
\quad
G  =  \frac{Y}{2(1+ \nu)},
\end{equation}

We have also calculated the BBSE corrections to the elastic moduli. These corrections modify the values of Young's moduli by maximally 10$\%$ and do not change the conclusion presented in this article.

At the end, we can compute the Poisson ratio values as follows $\nu  =  - (\Delta r / r)(l / \Delta l)$,
where $\Delta r$	 describes the change of the average radius of the functionalized CNT that has been caused by the applied strain $\Delta l$.

\section{\label{sec:res}Results and discussion}

\subsection{Influence of functionalization on the structure}

We have studied the stability and electronic structure of covalently functionalized CNTs in previous works \cite{app1, diamond, prof1, condmat}. We have also shown there how the functionalization induces changes in morphology of the functionalized systems and leads to the redistribution of electronic charge. All of the functionalizing fragments considered in the present study induce rehybridization from sp$^2$ to sp$^3$ of C-C bonds in neighborhood of the attachment, but in many cases we have found out that some of the adsorbed molecules also cause stronger deformation of CNT backbone structure. These pronounced changes in the morphology of the functionalized CNTs we observed motivated us to study the global strength of the functionalized CNTs expressed by the elastic moduli. 
Before we turn to the discussion of the elastic properties, we would like to present shortly the stability of the functionalized CNTs, and the change of geometry (lattice constant and radius) caused by the functionalization. 

\begin{figure} 
\includegraphics[width=0.45\textwidth]{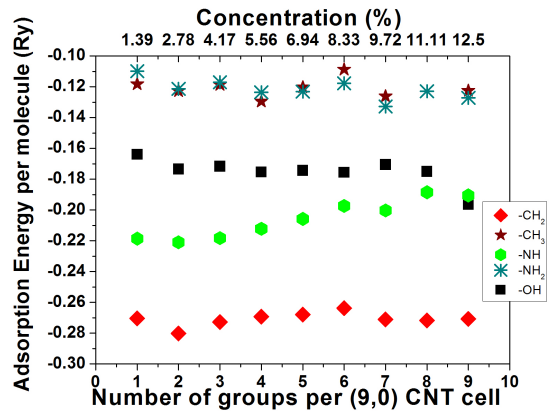}
\caption{\label{fig:Fig2}(color online) Adsorption energy per molecule of the (9,0) CNT functionalized with -NH, -NH$_2$, -CH$_2$, -CH$_3$, and -OH   groups as a function of the number of adsorbed fragments per CNT unit cell, i.e., per 72 carbon atoms. On the top axis, the universal percentage scale is depicted.}
\end{figure}

\begin{figure} 
\includegraphics[width=0.45\textwidth]{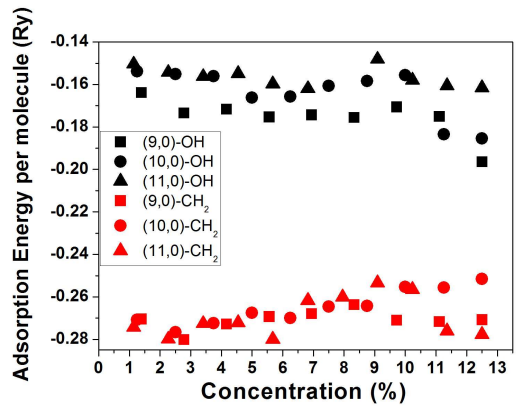}
\caption{\label{fig:Fig3} (color online) Adsorption energy per molecule for (9,0), (10,0), and (11,0) CNT functionalized with -CH$_2$  and -OH  fragments as a function of the density of attached molecules.}
\end{figure}

The adsorption energy (per adsorbed molecule) for all considered functionalizing molecules is shown in Fig.~\ref{fig:Fig2} for the prototypical metallic (9,0) CNT. It is seen that all the considered molecules bind to the surface of the (9,0) CNT (i.e., the adsorption energy is negative). However, the strength of the bonding is larger for typical radicals (-NH and -CH$_2$ ) then for the non-radicals (-NH$_2$ , -CH$_3$ ,  and -OH ). This we would like to correlate with the induced changes of geometry and elastic moduli later on. 
As can be seen in Fig. ~\ref{fig:Fig2}, generally, the adsorption energy per molecule remains nearly constant  with increasing number of attached molecules. Only for the strong radical -NH, the adsorption energy per molecule gets less negative (indicating that the bonding weakens) with increasing number of attached fragments. 
This trend obeys also for semiconducting (10,0) and (11,0) CNTs, as it has been illustrated for -CH$_2$  and -OH  adsorbents in Fig.~\ref{fig:Fig3}. At least for these rather similar in diameter CNTs and considered concentration of the adsorbed molecules, the adsorption energy depends rather weakly on the metallic or semiconducting character of the functionalized CNTs and their radius.   
\begin{figure} 
\includegraphics[width=0.45\textwidth]{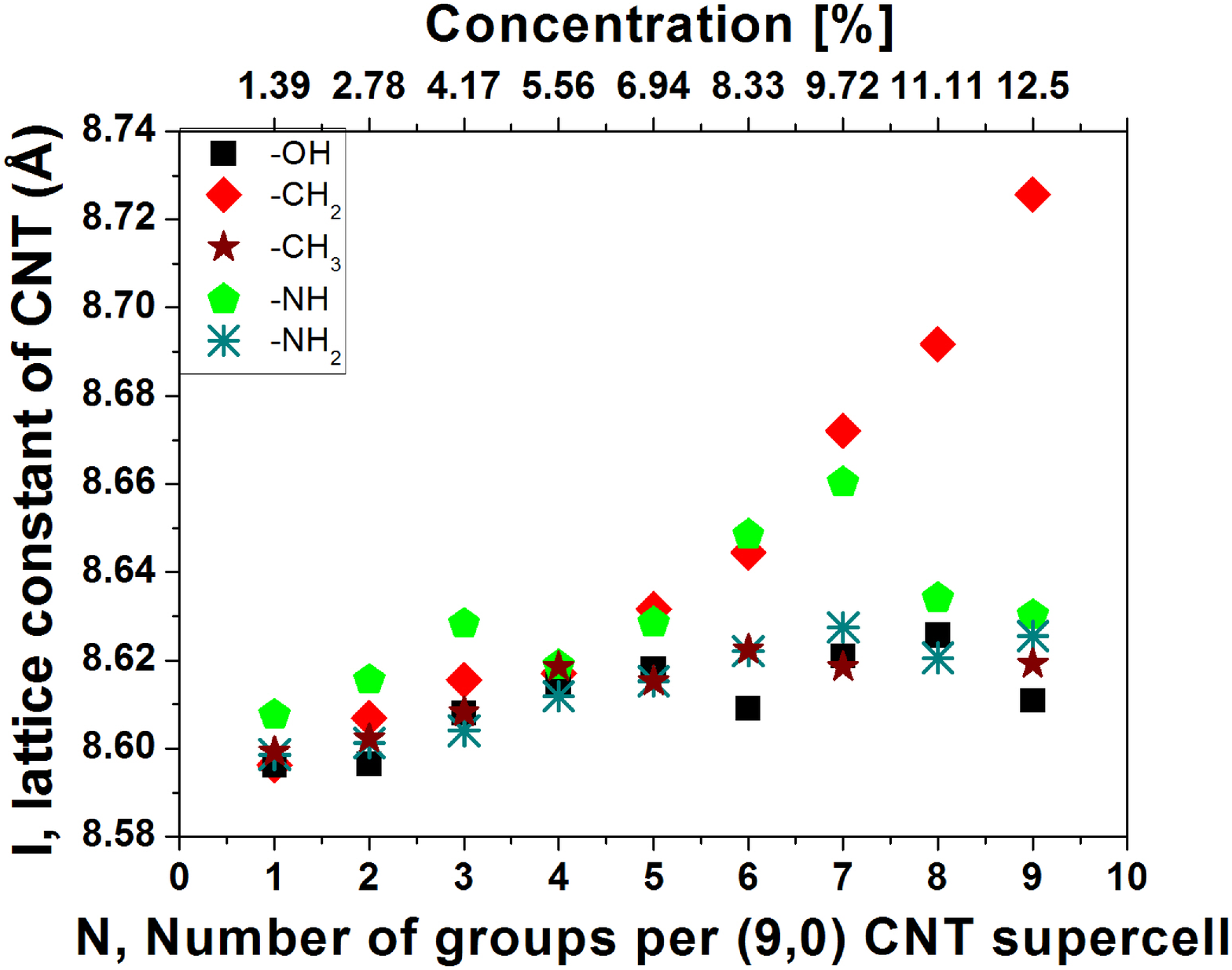}
\caption{\label{fig:Fig4} (color online) The equilibrium lattice constant ($l$) along symmetry axis of the functionalized nanotubes as a function of the number of covalently bound fragments to the sidewall of (9,0) CNTs for -NH, -NH$_2$, -CH$_2$, -CH$_3$, and -OH  functionalizing molecules. Top axis gives the concentrations of adsorbed molecules in \%.}
\end{figure}
\begin{figure} 
\includegraphics[width=0.45\textwidth]{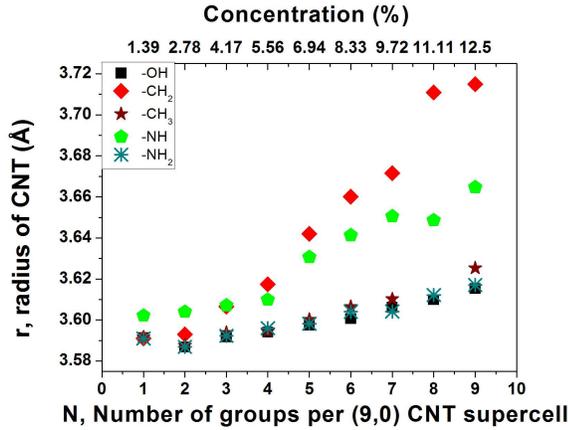}
\caption{\label{fig:Fig5} (color online) Radius of  (9,0) CNT functionalized with -NH, -NH$_2$, -CH$_2$, -CH$_3$, and -OH   groups as a function of number of covalently bound fragments to sidewall of the tube. Top axis gives the concentrations of adsorbed molecules in \%.}
\end{figure}
The functionalization changes the parameters characterizing the backbone of the functionalized CNTs, the longitudinal lattice constant $l$ and radius $r$. Longitudinal lattice constants and radii of the functionalized CNTs are larger than of pristine ones and change rather strongly with the number of attached molecules. It is depicted for the (9,0) CNT in Figs.~\ref{fig:Fig4} and ~\ref{fig:Fig5}. For example, radius and lattice constant of pure (9,0) CNT equals to 3.592 \AA $ $ and 8.590 \AA, respectively. For (9,0) CNT functionalized with 9 -CH$_2$  molecules per unit cell, the radius increases to 3.715 \AA $ $(by 3.31$\%$) and the lattice constant reaches  8.726 \AA $ $ (increase by 1.58$\%$).

Generally, one can say that functionalization of (9,0) CNT acts as an effective tensile strain, which blows up pristine CNT. The effect is more pronounced for molecules that built strong covalent bonds to the CNT walls.
The largest changes of the lattice constant $l$ and radius $r$ have been observed for CNT functionalized with -CH$_2$ radicals. For maximal considered concentration of 12.5\%, the relative changes of $l$ and $r$ in comparison to the length and radius of the pristine CNTs are equal to 1.56\% and 3.31\%, respectively. This effect is much weaker for -CH$_3$ functionalized CNT, where percentage change of $l$ is equal to 0.34\%, whereas the change of $r$ equals 0.92\%.

\begin{figure} 
\includegraphics[width=0.45\textwidth]{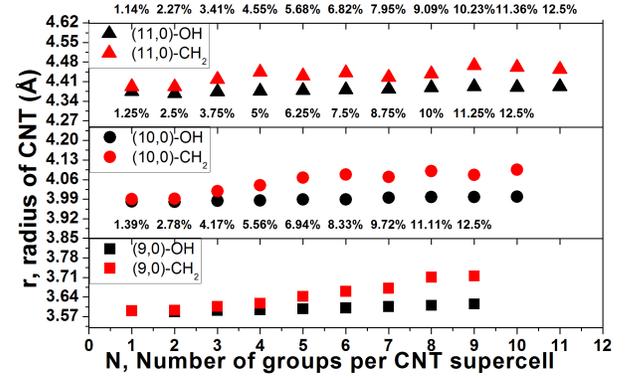}
\caption{\label{fig:Fig6} (color online) Radius of (9,0), (10,0), and (11,0) CNTs functionalized with -CH$_2$  and -OH  as a function of the number of attached fragments per supercell. Since (9,0), (10,0), and (11,0) CNTs contain different number of atoms in the supercell, the percentage concentrations of the attached fragments are also depicted above top axes of each panel for better comparison.}
\end{figure}

The relative changes in the $l$ and $r$ parameters induced by functionalization depend rather slowly on metallic - semiconducting character of CNTs and their diameter. The radius of (9,0), (10,0), and (11,0) CNTs functionalized with -CH$_2$  and -OH  molecules is depicted in Fig.~\ref{fig:Fig5}. As it was determined previously\cite{condmat} -CH$_2$  radicals bind strongly to the CNT surfaces and at higher concentrations can lead to some local structural defects (so-called 5-7 defects). On the other hand, the functionalization with -OH  groups slightly changes the cross-section of CNT - from circle to ellipse. Therefore, we have decided to compare (see Fig.~\ref{fig:Fig6}) both types of attachments for all of the CNTs studied: (9,0), (10,0) and (11,0). We have noticed, for the biggest considered concentration of 12.5$\%$, that (9,0) CNT functionalized with -CH$_2$  shows the biggest percentage change ( 3.31$\%$) of radius in comparison to (10,0) and (11,0) CNTs (where percentage changes are equal to 2.83$\%$ and 1.81$\%$, respectively). The -OH  groups follow the similar trend, however, the functionalization induced changes of the radius are weaker. The relative changes of the radius are 0.64$\%$, 0.43$\%$ and 0.39$\%$,  for (9,0), (10,0) and (11,0) CNT, respectively.  Therefore, one can say that the functionalization of the nanotubes with larger original radius has less influence on its structure than functionalization of CNTs with smaller diameter.  

Having described the equilibrium geometry of the functionalized CNTs, we are now in the position to discuss their elastic properties. 

\subsection{Elastic properties of pure CNT}

Before we turn to the elastic moduli of the functionalized CNTs, we would like to present our results for pristine (9,0), (10,0), and (11,0) ones. This allows for comparison with previous works and provides the reference to the case with functionalization.

\begin{table}[h!tb] \centering
\caption{\small{Elastic moduli and Poisson's ratio of (9,0), (10,0), and (11,0) pristine CNTs.}}
\vspace{0.06 in}
\begin{tabular}{|l|c|c|c|}
\hline
    \small{\textbf{Property}} & \small{\textbf{(9,0)}} &   \small{\textbf{(10,0)}} &  \small{ \textbf{(11,0)}}  \\
\hline
\hline
Y (TPa)   &     1.02 &    1.03 &     1.02  \\
\hline
K (TPa) &     0.61   &  0.57   &    0.54     \\
\hline
G (TPa)  &  0.41    &   0.43   &  0.43     \\
\hline
$\nu$ &   0.22     &    0.20       &    0.18        \\
\hline
\end{tabular}  
\label{tab:tab1} 
\end{table}  

Young's, Shear, and Bulk moduli, and also Poisson's ratios for (9,0), (10,0) and (11,0) pristine CNTs are gathered in Tab.\ref{tab:tab1}. The calculated Young's moduli of the pure CNT compare excellently to experimental findings (0.32-1.80 TPa) \cite{lu1997, krishna1998, Terrones2003} and previous theoretical works (0.8-1.5 TPa) \cite{lier2000, li2003, hernandez1998, govindjee1999, chang2006, yao1998, xin2000, Terrones2003, kudin2001, sanchez1999}. Calculated Poisson's ratios are identical to the experimental ones and also very close to previous theoretical  predictions  (0.19-0.34) \cite{lu1997, popov2000, chang2006, Terrones2003, sanchez1999}. Also calculated values of the Shear and Bulk moduli agree fairly well with previously obtained theoretical and experimental values, lying in the range of 0.45-0.58 TPa \cite{li2003, Terrones2003, krishna1998, lu1997, chang2006, popov2000} and 0.50-0.78 TPa \cite{Terrones2003, lu1997}, respectively.

We have also calculated the elastic properties for wider range of zigzag pristine CNTs. Only for small CNTs, like (4,0) and (5,0) all the values of elastic moduli are smaller.
  For larger in diameter CNTs, up to (20,0), the values are very similar to those shown in Tab.\ref{tab:tab1}. Starting from (6,0) CNT, all of the elastic moduli seem to be rather weakly dependent on the diameter of CNT. Such behavior of Young's as well as Shear modulus has been noticed in previous studies for Young's\cite{popov2000, chang2006,  sanchez1999, li2003, shokrieh2010, lu1997} and for Shear\cite{popov2000, li2003, lu1997} moduli.

\subsection{Elastic properties of functionalized CNT}

\begin{figure} 
\includegraphics[width=0.45\textwidth]{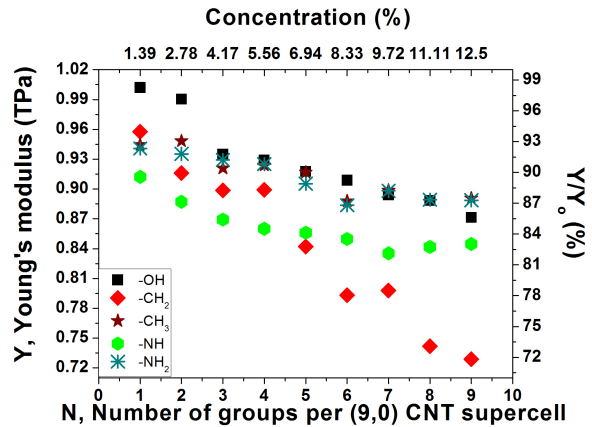}
\caption{\label{fig:Fig7}(color online) The Young's modulus of the (9,0) CNT functionalized with -NH, -NH$_2$, -CH$_2$, -CH$_3$, and -OH  groups as a function of the density of attached molecules, given as the number of attached molecules per unit cell (lower x-axis) or the ratio of adsorbents to the number of atoms in the unit cell (upper x-axis). On the right axis we have depicted percentage change of Young's modulus relative to the pristine CNT.}
\end{figure}

Let us now present theoretical predictions for elastic moduli of the functionalized CNTs. We start the presentation of our results with Young's modulus of the (9,0) CNT functionalized with -NH, -NH$_2$, -CH$_2$, -CH$_3$, and -OH groups (Fig.~\ref{fig:Fig7}). For all considered groups,  the Young's modulus decreases with increasing density of the attachments. However, for the radicals, -CH$_2$  and -NH, the trend is much more pronounced than for other groups. For CNTs functionalized with 9 -CH$_2$  (i.e., concentration of 12.5\%), the Young's modulus decreases by 28.41\%, whereas CNT with 9 functionalizing -CH$_3$  groups exhibits reduction in the Young's modulus equal to 13.52\%. It confirms the already described tendency that the molecules with stronger binding to the CNT's surface modify the properties of the functionalized CNTs in a stronger manner. 

\begin{figure} 
\includegraphics[width=0.45\textwidth]{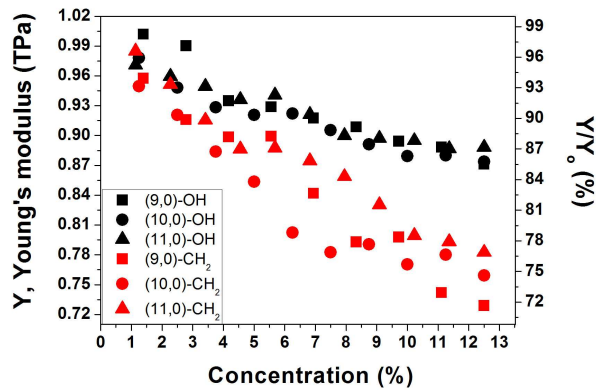}
\caption{\label{fig:Fig8}(color online) The Young's modulus of (9,0), (10,0), and (11,0) CNTs functionalized with -OH  and -CH$_2$  fragments as a function of the density of attached molecules.}
\end{figure}

For the purpose of comparison how the Young's modulus depends on diameter of the tubes, we have chosen -OH  groups and -CH$_2$  fragments  as examples.    
In Fig.~\ref{fig:Fig8}, we plot the dependence of Young's modulus for (9,0), (10,0) and (11,0) CNTs functionalized with -OH  and -CH$_2$  molecules on the density of adsorbents. It is seen that -OH  groups represent behavior typical for non-radical adsorbents (which generally cause small deformation of CNTs), and one observes practically no difference between tubes. Even in the case of -CH$_2$  radical (that causes typically rather large deformations of CNTs), one can only weakly differentiate between the types of the functionalized nanotubes. 

Our calculations show that Poisson ratio for structures functionalized by all considered fragments always oscillates between values 0.17 and 0.24. This quantity, for the studied range of the adsorbent concentrations, neither exhibits clear dependencies on the type of functionalizing molecules nor allows for resolution between (9,0), (10,0), and (11,0) CNTs.

We complete the discussion of the elastic moduli for functionalized CNTs with the presentation of results for Shear and Bulk moduli, the magnitude of which can be easily calculated from Young's modulus and Poisson ratio employing formulas \ref{e5}.

\begin{figure} 
\includegraphics[width=0.45\textwidth]{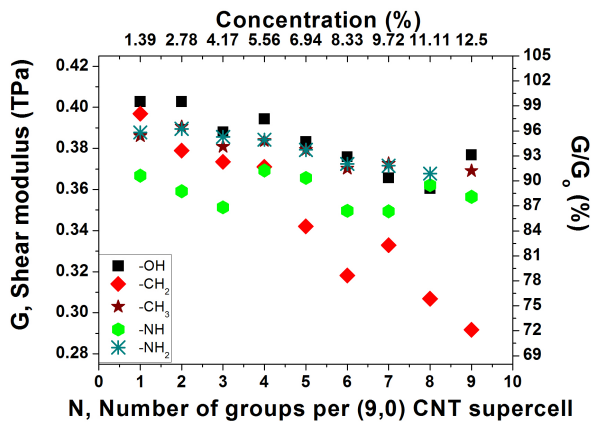}
\caption{\label{fig:Fig9} (color online) The Shear modulus of (9,0) CNT functionalized with -NH, -NH$_2$, -CH$_2$, -CH$_3$, and -OH  groups as a function of the density of attached molecules, given as the number of attached molecules per unit cell (lower x-axis) or the ratio of adsorbents to the number of atoms in the unit cell (upper x-axis). On the right axis we have depicted the relative changes of Shear modulus with respect to pristine CNT.}
\end{figure}
The Shear modulus as a function of the concentration of attached molecules is depicted in Fig.~\ref{fig:Fig9}. Generally, the Shear modulus drops with the increasing density of the attached molecules. This decrease is stronger for -CH$_2$  radical than for non-radical groups such as -OH .  
For the highest considered concentration of the -CH$_2$  radicals, the Shear modulus is smaller by roughly 25\%, and even for the non-radical functionalizing groups the decrease is of the order of 10\%. Therefore, our studies do not corroborate Franklad's \cite{frankland2002} suggestion that functionalization has tiny influence on Shear modulus (less then 4.63 $\%$).

\begin{figure} 
\includegraphics[width=0.45\textwidth]{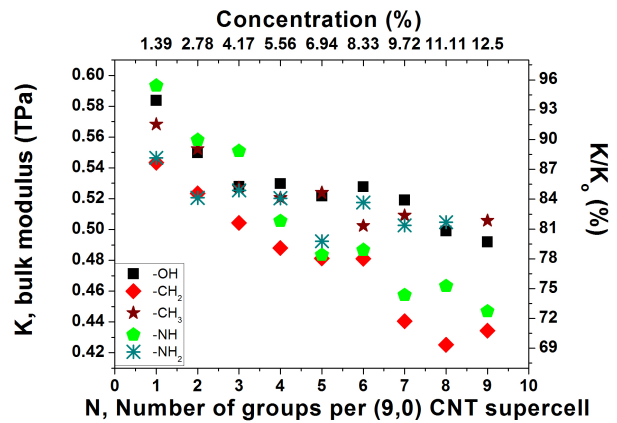}
\caption{\label{fig:Fig10}(color online) The Bulk modulus of (9,0) CNT functionalized with -NH, -NH$_2$, -CH$_2$, -CH$_3$,  and -OH  groups as a function of the density of attached molecules, given as the number of attached molecules per unit cell (lower x-axis) or the ratio of adsorbed molecules to the number of atoms in the unit cell (upper x-axis). On the right axis we have depicted percentage change of the Bulk modulus with respect to pristine CNT.}
\end{figure}

The Bulk modulus as a function of the concentration of attached molecules is shown in Fig.~\ref{fig:Fig10}. 
The Bulk modulus behaves similarly to other elastic moduli and decreases with the growing concentration of functionalizing molecules, with the strongest effect observed for functionalization with -CH$_2$  radical. 

Generally, our studies provide theoretical predictions for the elastic moduli of the covalently functionalized CNTs. These moduli diminish with the concentration of the functionalizing molecules. In the situation of the lack of experimental data, the obtained values should facilitate the undestanding and design of the composite materials. First of all, the decrease of elastic moduli is quite modest, particularly for non-radical -OH, -NH$_2$, -CH$_3$  groups. Therefore, the functionalized CNTs should still be good reinforcement in composites employing polymers or metals as matrices. On the other hand, the functionalization of CNTs is necessary to bind CNTs to polymer matrix and significantly improves homogeneous dispersion and integration of CNTs into polymers, simultaneously reducing the tendency of pristine CNTs to re-agglomeration. This feature substantially enhances elastic strength of polymer matrices with incorporated CNTs. This effect has been confirmed in a series of experiments studying the Young's modulus of composites with amines \cite{gojny2005, wang, lachman2010}, amides \cite{steiner2012}, and carboxylic groups \cite{lachman2010, amr2011}. All these studies are in agreement with our findings. However, for CNT with hydroxyl groups dispersed into polymer matrix, Wang \cite{wang2011} reported reduction in Young's modulus in comparison to pure matrix. Unfortunately, the issue of the elastic properties of composites is out of scope of our atomistic approach, and would require a study based on continuous methods. 

\section{\label{sec:con}Conclusions}

We have performed extensive and systematic {\sl ab initio} studies of the elastic properties of the (9,0), (10,0), and (11,0) CNTs functionalized with -NH, -NH$_2$, -CH$_2$, -CH$_3$, -OH  molecules covalently bound to the CNT walls at concentrations reaching up to 4.6$\cdot$10$^{14}$ molecules per cm$^2$. Our studies provide valuable  theoretical quantitative predictions for elastic moduli (Young's , Shear, Bulk moduli, and Poisson ratio) of functionalized CNTs, demonstrate clear chemical trends in the elastic moduli, and shed light on physical mechanisms governing these trends. These results are of importance for design of composite materials employing carbon nanotubes.

We have shown that considered molecules form covalent bonds to the CNT surfaces and cause local and global changes in the morphology of the CNT that are generally proportional to the density of the attached molecules. 
The local deformations include rehybridization of the C-C bonds and defects that influence strength of the functionalized systems. Functionalization of CNTs causes expansion of the functionalized CNTs, i.e., increase of longitudinal lattice constant and radius in comparison to the pristine CNTs. This expansion is proportional to the density of the adsorbed molecules. 
We observe general trend that the molecules forming the stronger bonds to CNTs cause larger deformations of the functionalized systems (i.e., the larger changes of the lattice constants $l$  and radii $r$) and larger reduction of the elastic moduli (Young's, Shear, and Bulk). All moduli decrease with concentration of the adsorbed molecules. 
As far as the Young's, Shear, and Bulk moduli reflect changes in the CNT morphology caused by functionalization, the Poisson's ratio remains almost unchanged.
In a few cases when comparison with experimental or other theoretical studies is possible, we observe reasonable agreement with results of our calculations. 
In spite of the fact that the functionalization diminishes elastic moduli of CNTs and this effect generally cannot be neglected, the elastic moduli remain large enough to guarantee successful employment of functionalized CNTs for reinforcement of composite materials.

\section{Acknowledgement}

The authors gratefully acknowledge financial support of the Polish Council for Science through the Development Grants for the years 2008-2011 (NR. 15-0011-04/2008, NR. KB/72/13447/IT1-B/U/08) and the SiCMAT Project financed under the European Founds for Regional Development (Contract No. UDA-POIG.01.03.01-14-155/09).  We thank also PL-Grid Infrastructure and Interdisciplinary Centre for Mathematical and Computational Modeling of University of Warsaw (Grant No. G47-5) for providing computer facilities. 

\end{document}